\documentclass[aps,prl,showpacs,amsfonts,amssymb,amsmath,floatfix,twocolumn,
floats,nobalancelastpage]{revtex4}

\usepackage{graphicx}

\begin{document}          

\title{Subclasses in Mixing Correlation-Growth Processes with Randomness}

\author{A. Kolakowska}
\author{M. A. Novotny}
\affiliation{Department of Physics and Astronomy, and Center for 
Computational Sciences, 
P.O. Box 5167, Mississippi State, MS 39762-5167}

\date{\today}

\begin{abstract}
We show that in the construction of continuum equations for competitive  
growth processes that are a mixture of random deposition and a correlation process, 
a distinction must be made within a \textit{single universality class} between 
depositions that do and do not create voids in the bulk. Within these 
subclasses the bulk morphology is reflected in the surface roughening 
via \textit{nonuniversal} prefactors in the universal scaling of the surface width. 
\end{abstract}

\pacs{81.15.Aa, 02.50.Fz, 68.55.Ac}

\maketitle

Many complex systems are studied by mapping onto a suitable nonequilibrium 
surface-growth problem \cite{mix,KN05}. The dynamics of the buildup of the 
correlations in a system can be then explored with surface-growth methodologies. 
Large-scale properties are described within a continuum model by 
universal stochastic growth equations and tested with simulation models. 
The trouble is, simple atomistic models are often not adequate to reproduce the 
complex physics of the observed surface phenomena that may involve 
contributions from several universal processes, but the continuum description 
of such multicomponent growth has not yet been developed. A representative 
example comes from an applied model in computer science \cite{KN05} when 
the asynchronous dynamics of conservative updates in a system of parallel processors 
is modeled as a virtual-time surface that represents nonequilibrium processes 
in this system. When the load per processor is minimal, this dynamics is 
Kardar-Parisi-Zhang (KPZ) universal process \cite{KTN+00}. But when the load 
is increased to reflect real operations, the realistic dynamics is a competitive 
growth that combines a universal KPZ process with a random 
deposition (RD) process, i.e., is of the type RD+KPZ \cite{KNV04}. 
Consequently, it is important to know how 
the \textit{nonuniversal} properties of 
the component-processes affect the universal scaling of a RD+KPZ process. 
This is a still unsolved problem of nonequilibrium surface growth science. 
In applied modeling, even if not explicitly assumed, competitive dynamics 
naturally arises. Studying these systems should also contribute to the understanding   
of differences between the expected and the actual scaling of rough interfaces, 
often encountered both in simulations and in experiments.

In this work we investigate a connection between surface roughening 
and the bulk morphology formed during the deposition in competitive growth 
RD+X, where X is a correlation growth process of universal dynamics different  
from RD. This connection has been already established in simulations of 
competitive growth models \cite{PJ90} and binary growth of thin 
films \cite{DK00}. A new aspect of our study is a direct 
theoretical link between nonuniversal properties of process X, as read 
from the bulk, and the continuum equation that reflects the observed 
scaling laws for the competitive RD+X processes. 
We derive from first principles a continuum equation 
to show that model dependent coefficients do reflect the bulk structure. 
This will lead to a distinction between void-producing and desorption 
growth processes and simple absorption processes. As discussed later, 
this division into subclasses is a {\it necessary} first step towards a  
theory of many-component processes. In particular, 
it explains variations in scale-dilatation observed in RD+X 
models \cite{KNV04,KNV06,Reis06,HA06,HA03,HMA01,HA01}. 
In our analysis we use as an example universal RD+KPZ  growth processes 
in $(1+1)$ dimensions, and generalize our approach to other processes 
in $(1+n)$ dimensions.

A scaling hypothesis for competitive RD+X processes  \cite{KNV06} 
states that if a correlation growth X occurs with a constant probability 
$p$, its continuum equation must be invariant under the scaling
\begin{equation}
\label{scaling}
x \to x \, , \; h \to h/g(p) \, , \; t \to t/f(p) \, ,
\end{equation}
where $g(p)$ and $f(p)$ are arbitrary suitable functions of $p \in (0;1]$. 
This invariance implies that $f(p)=g^2(p)$, and when X=KPZ 
it leads to the KPZ equation \cite{KPZ86} 
for the RD+KPZ mix \cite{KNV06}:
\begin{equation}
\label{data}
h_t = \nu_0 f(p) h_{xx} + (\lambda_0 /2) f^{3/2}(p) h_x^2 + \eta (x,t) \, ,
\end{equation}
where $h \equiv h(x,t)$ is the height field; $x$ and $t$ are the spacial 
and time coordinates, respectively; subscripts denote partial 
derivatives; $\eta (x,t)$ is the white noise; and,  
$\nu_0$ and $\lambda_0$ are constants. When $\lambda_0=0$, Eq.~(\ref{data}) is 
the Edwards-Wilkinson (EW) equation \cite{EW82} when X=EW. 
When $\nu_0=\lambda_0=0$, Eq.~(\ref{data}) defines universal 
RD dynamics. Many simulation models of RD+EW and RD+KPZ growth 
processes \cite{KNV06,HA03} suggest $g(p)=p^\delta$ in 
Eq.(\ref{scaling}), which leads to the Family-Vicsek universal 
scaling \cite{FV85} of the average surface width $w(p,t)$ \cite{KNV06}: 
\begin{equation}
\label{FV}
w(p,t) = \frac{L^\alpha}{p^\delta} F \left( p^{2\delta} \frac{t}{L^z}\right) \, .
\end{equation}
For substrates of size $L$, $F(y)$ describes two limit-regimes of evolution: 
$F(y) \sim y^{\alpha /z}$ if $y \ll 1$ (growth); and, $F(y) \sim \textrm{const}$ 
if $y \gg 1$ (saturation). In Eq.~(\ref{FV}), $\alpha$ and $z$ are the universal 
roughness and dynamic exponents, respectively, of the universality class 
of the correlation growth X. The scale-dilatation exponent $\delta$ in scaling  
prefactors in Eq.(\ref{FV}), however, is {\it nonuniversal}. It has been observed 
that in some models $\delta \approx 1$ across universality classes, and in  
some other models $0<\delta \lessapprox 1$ within a single universality 
class \cite{KNV04,KNV06,Reis06}. Also, there are 
models where prefactors in Eq.(\ref{FV}) do not at all obey a power 
law in $p$ \cite{note02}. Here, we shall establish that this variation is 
not accidental, but rather reflects the properties of the bulk of the 
deposited material.

Consider aggregations where particles fall onto a  
substrate of $L$ sites, where they may be accepted in accordance 
to a rule that generates correlations among the sites. This correlation  
growth occurs with probability $p$ and competes with RD growth that 
occurs with probability $q=1-p$. When a particle is accepted at a site, 
the site increases its height by $\Delta h$. If, e.g., component {\it 1} is 
RD, and component {\it 2} is a correlation growth in the KPZ 
universality class, their corresponding growth equations are
\begin{eqnarray}
h_{1, t} &=& \eta_1 (x ,t) \, ,  \label{comp-1} \\
h_{2, t} &=& \nu_0 h_{2, xx} + (\lambda_0 /2) h_{2, x} ^2 + 
\eta_2 (x,t) \, ,  \label{comp-2} 
\end{eqnarray}
where $h_n(x,t)$, $n=1,2$, is the column height at $x$ after time $t$ 
when the component {\it n} acts alone. Assume for simplicity that the 
noise terms are of the same strength, i.e., $\eta \equiv \eta_1 = \eta_2$. 
In two-component growth, when both components act simultaneously together, 
the column height $h(x,t)$ is incremented due to either of the components 
with their corresponding probabilities $\tilde{p}$ 
and $\tilde{q}$, $\tilde{p}+\tilde{q}=1$:
\begin{equation}
\label{simplex}
\Delta h(x,t) = \tilde{p} \Delta h_2 (x,t) + \tilde{q} \Delta h_1 (x,t) \, .
\end{equation}
Here, probability $\tilde{p}$ (or $\tilde{q}$) is the fraction of contributions 
to $h$ from component {\it 2} (or {\it 1}). For some processes this fraction 
is identical to a fraction of times when $h(x)$ is incremented due to 
component {\it 2} (or {\it 1}) for the times from $0$ to $t$. However, 
as explained later, this is not so for all processes. 
In Eq.~(\ref{simplex}), $\Delta h_n$ is understood as 
``being incremented due to the process {\it n},'' $n=1,2$. In this 
statistical sense, Eq.~(\ref{simplex}) expresses a simplectic 
decomposition of $\Delta h(x,t)$ into its vertex-components 
$\Delta h_n(x,t)$. Dividing Eq.~(\ref{simplex}) by $\Delta t$, 
and taking the limit $\Delta t \to 0$, gives the equation for 
time rates, $h_t = \tilde{p} h_{2,t} + \tilde{q} h_{1,t}$, 
to which we substitute Eqs.~(\ref{comp-1})-(\ref{comp-2}): 
\begin{equation}
\label{combine}
h_t = \nu_0 \tilde{p} h_{2,xx} + (\lambda_0/2) \tilde{p} h_{2,x}^2
+ (\tilde{p}+\tilde{q}) \eta (x,t) \, .
\end{equation}
In Eq.(\ref{combine}), $h(x,t)$ is the column height that rises at $x$ as the result 
of two processes acting simultaneously from the beginning 
to time $t$. Here, $h_2(x,t)$ is the part of $h(x,t)$ that was 
created by the component {\it 2} in this time. The other part was 
created by component {\it 1}. In other words, $h_2(x,t)$ is so 
far an unknown fraction of $h(x,t)$. To find a 
relation between $h$ and $h_2$, one must consider nonuniversal 
properties of aggregation processes.

We distinguish between the following two groups of surface growths. 
In one group we place all simple absorption processes with conserved flux 
that do not create voids in the bulk of the deposited material. We call this group 
{\it absorption-bulk-compact} (ABC) growths. The other group, 
which we call {\it dense-or-lace-bulk} (DOLB) growths, contains 
processes that are not ABC-type. The DOLB group includes desorption 
processes that may lead to a dense bulk as well as absorptions 
that lead to the formation of voids. Note, RD processes are  
ABC growth processes. As we show in the next paragraph, when component {\it 2} 
is of the ABC-type, $\tilde{p}$ and $\tilde{q}$ in Eq.~(\ref{simplex}) express fractional 
contributions to $h$ in terms of times, and then $h_2(x,t)=ph(x,t)$. 
This is not true when component {\it 2} is a DOLB growth.

Consider a discrete representation of events at coordinate $x$. 
Suppose, there are $t$ deposition events in total, with $t_1$ 
events due to component {\it 1}, and $t_2$ events due to component 
{\it 2}, $t=t_1+t_2$. In ABC growth, after $t$ events, the total 
column height is $h=t \Delta h$, where contributions from components 
{\it 1} and {\it 2} are, respectively, $h_1=t_1 \Delta h$ and 
$h_2=t_2 \Delta h$. Thus, $h_1/h=t_1/t=q$ and $h_2/h=t_2/t=p$. 
Therefore, in ABC growth $h_2=ph$, and in 
Eqs.~(\ref{simplex})-(\ref{combine}) $\tilde{p}=p$ and $\tilde{q}=q$.

Next consider that the component {\it 2} is a DOLB growth that creates voids. 
Now, an individual deposition event due to component {\it 2} not 
only increases $h$ by $\Delta h$, but may also result in the 
creation of voids. The net effect is as though component {\it 2} 
deposited $\Delta h$ {\it and} the voids. Therefore, in $t_2$ events, 
its contribution to the column height is $h_2=(t_2+m) \Delta h$, 
where $m \Delta h$ reflects the increase in height due to the 
presence of voids. The component {\it 1} is RD, i.e., ABC-type, and 
$h_1=t_1 \Delta h$. After $t$ events, the net column height 
is $h=h_1+h_2=(t+m) \Delta h$. Thus, $h_1/h=t_1/(t+m) < t_1/t =q$ 
and $h_2/h=(t_2+m)/(t+m) > t_2/t = p$. Fractions 
$q_{\mathrm{eff}} \equiv h_1/h$ and $p_{\mathrm{eff}} \equiv h_2/h$ are 
the {\it effective probabilities} of deposition events due 
to components {\it 1} and {\it 2}, respectively, as they 
would result from measuring the column height. 
For some types of two-component growth with RD, the 
probability $p_{\mathrm{eff}}$ can be expressed approximately 
as $p_{\mathrm{eff}}=p^\delta$ \cite{note02}.  
For DOLB growths with voids $\delta <1$ (because $p_{\mathrm{eff}} > p$). 
When the component {\it 2} is a DOLB growth with desorption, in the 
above reasoning one should change $m \to -m$. This will give 
$q_{\mathrm{eff}} > q$ and $p_{\mathrm{eff}}<p$, and 
$p_{\mathrm{eff}}=p^\delta$ with $\delta >1$. 
The value of $\delta$ is specific to the particular deposition process of 
component {\it 2}. Therefore, in DOLB growth $h_2=p_{\mathrm{eff}}h$, 
and in Eqs.~(\ref{simplex})-(\ref{combine}) 
$\tilde{p}=p_{\mathrm{eff}}$ and $\tilde{q}=q_{\mathrm{eff}}$.

In general, $h_2 (x,t)= p_{\mathrm{eff}}h(x,t)$ and 
$\tilde{p} \equiv p_{\mathrm{eff}}$, where $p_{\mathrm{eff}}=p$ if 
the correlation component is an ABC growth. When the correlation component is 
a DOLB growth, and when the effective probability is well approximated by 
a power law $p^\delta$, the result can be summarized as $p_{\mathrm{eff}}=p^\delta$, 
where $\delta=1$ for ABC growths and 
$\delta \ne 1$ for DOLB growths. This result is combined with 
Eq.~(\ref{combine}) to give the continuum equation for the RD+KPZ mix:
\begin{equation}
\label{RD_KPZ}
h_t=\nu_0 p^{2\delta} h_{xx} + (\lambda_0 /2) p^{3\delta} h_x^2 + \eta (x,t) \, .
\end{equation}
When in Eq.~(\ref{comp-2}) $\lambda _0 \equiv 0$, the analogous 
reasoning gives the RD+EW dynamic:
\begin{equation}
\label{RD_EW}
h_t=\nu_0 p^{2\delta} h_{xx} + \eta (x,t) \, .
\end{equation}

\begin{figure}[tp]
\includegraphics[width=7.5cm]{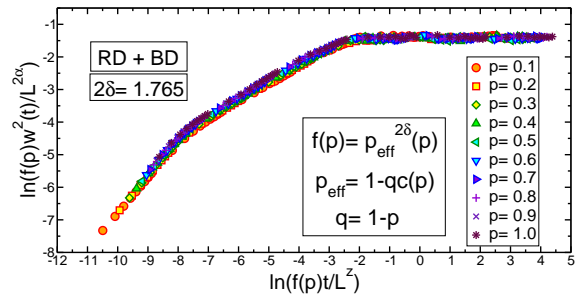}
\caption{\label{effective_scaling} 
(color on line) 
Scaled time-evolution $w^2(p,t)$ in the RD+BD model. In this example,
the scaling function $f(p)=g^2(p)$ explicitly incorporates the compactness
$c(p)$ of the bulk formed in the RD+BD process. Here, $L=500$,
$2\alpha=1$, and averaging was performed over 400 configurations.
}
\end{figure}

Both results, Eqs.(\ref{RD_KPZ})-(\ref{RD_EW}), are in accord with 
our former derivation  that lead to Eq.(\ref{data}) \cite{KNV06}. 
Matching Eq.~(\ref{RD_KPZ}) with Eq.~(\ref{data})  
gives $f(p)=p^{2\delta}$, which form of $f(p)$ was used formerly  
to derive the approximate prefactors in Eq.(\ref{FV}). 
The inverse of the scaling (\ref{scaling}) when applied to 
Eqs.(\ref{RD_KPZ})-(\ref{RD_EW}) transforms them to 
continuum equations for a ``pure'' correlation processes of $p=1$. 
Explicitly, it collapses {\it all} evolution curves $w(p,t)$ 
(for all $L$ and $p$) either onto $w(1,t)$ or onto a neighborhood 
of $w(1,t)$ \cite{KNV06}, following Eq.(\ref{FV}), 
provided the effective probabilities $p_{\mathrm{eff}}$ can be 
approximated by the power-law $p^{\delta}$. When such a fit is not 
possible Eq.(\ref{FV}) is obeyed but then the scaling prefactors 
must be expressed directly in terms of effective probabilities. 
This is because the factor $p^\delta$ in the coefficients of 
Eqs.(\ref{RD_KPZ})-(\ref{RD_EW}) is only a fit to the 
effective probability $p_{\mathrm{eff}}$. In fig.\ref{effective_scaling} 
we give an example of the exact scaling where nonuniversal 
prefactors in Eq.(\ref{FV}) are directly expressed by $p_{\mathrm{eff}}$ 
via the substitution $p^{\delta} \to p_{\mathrm{eff}}^{\delta}(p) = \sqrt{f(p)}$ 
for the RD+BD model when BD is the NN sticking 
rule \cite{BS95}. Here, the effective probability depends on both $p$ and 
the mean compactness $c(p)$ of the bulk formed in the RD+BD process: 
$p_{\mathrm{eff}}=1-qc(p)$ \cite{note02}. The perfect data collapse in the 
\textit{full range} of $p\in (0;1]$, seen in fig.\ref{effective_scaling}, 
can be contrasted with fig.5 of Ref.\cite{KNV06} that shows only an 
approximate data collapse for the same system with the best fit exponent 
$\delta \approx 0.41$ in Eq.(\ref{FV}). It needs to be said explicitly 
that the scaling where $\delta=1/2$ in Eq.(\ref{FV}), proposed in 
Refs.\cite{HA06,BL05} for RD+BD models, does not produce data 
collapse at all. 
The RD+BD model when BD is the NNN sticking rule \cite{BS95} 
provides an example where $p_{\mathrm{eff}}(p)$, 
and thus the nonuniversal prefactors $f(p)$ and $g(p)$  
in Family-Vicsek universal scaling, cannot be expressed by 
a power-law $p^\delta$. In this system the surface roughening 
obeys power laws in effective probability that incorporates either 
the compactness or the voidness of the bulk, which gives excellent 
data collapse of $w(p,t)$, similar to that 
seen in fig.\ref{effective_scaling} \cite{note02}.

The approach introduced here by the example of KPZ processes, 
can be applied to a broad range of stochastic growth models RD+X, where 
component {\it 2} can be any isotropic growth in $(1+n)$ dimensions:
\begin{equation}
\label{process-2}
h_{2,t}(\vec{x},t)=G(h_2)  + \eta_2 (\vec{x},t) \, ,
\end{equation}
where $\vec{x}$ is $n$ dimensional, and the operator $G$ represents 
only local interactions \cite{BS95}. 
In the general case, Eq.(\ref{combine}) is written as  
$h_t=p_{\mathrm{eff}} h_{2,t} + q_{\mathrm{eff}} h_{1,t}$, and  
combined with Eqs.(\ref{comp-1}) and 
(\ref{process-2}), to find for the competitive growth
\begin{equation}
\label{competitive}
h_t (\vec{x},t) = p_{\mathrm{eff}} G(p_{\mathrm{eff}} h)  + \eta (\vec{x},t) \, ,
\end{equation}
where $\eta = (1-p_{\mathrm{eff}}) \eta_1 + p_{\mathrm{eff}} \eta_2$, 
and the noise strengths may be different.   
Eqs.(\ref{process-2})-(\ref{competitive}) represent the same  
universality class since the multiplication by  $p_{\mathrm{eff}}$ 
does not modify local interactions:  $p_{\mathrm{eff}}$ affects the noise 
strength and the gradient of the height field, but it does not generate new terms 
other than those already given by operator $G$. Hence, if a correlation growth 
belongs to a given universality class, its mix with RD will remain in the same class. 
Elementary calculations show that Eq.(\ref{competitive}) is invariant 
under the scaling $g(p) h(\vec{x},t)=h'(\vec{x},t'=f(p)t)$. 
If $g(p)=p_{\mathrm{eff}}(p)$ and $f(p)=p_{\mathrm{eff}}^2 (p)$, and if 
the noise strengths are the same, this scaling maps the universal dynamics 
(\ref{competitive}) of RD+X onto the universal dynamics of X. 
In this case the invariance implies $g(p)w(p,t)=w'(f(p)t)$, 
where $w'(\cdot)$ has universal scaling properties of the process X. 
When X is either in the KPZ or in the EW universality 
class, and if additionally $p_{\mathrm{eff}}\approx p^{\delta}$, we 
recover Eq.(\ref{FV}).

When both the RD and the correlation component {\it 2} 
have deposits of unit height, when $p_{\mathrm{eff}}\approx p^{\delta}$, 
we have $\delta =1$ if component {\it 2} is of the ABC-type; and, $\delta \ne 1$ if it 
is of the DOLB-type. In the latter case, the value of the exponent 
$\delta$ is specific to component {\it 2}. 
When $p_{\mathrm{eff}}$ incorporates explicitly bulk properties, 
the scaling is $g(p)=p^{\delta}_{\mathrm{eff}}(p)$, 
where the new scale-dilatation exponent $\delta$ is obtained from 
the slope of $\ln{w^2(p)}$ plotted vs $\ln{p_{\mathrm{eff}}(p)}$ at saturation. 
In DOLB growth with voids, $p_{\mathrm{eff}}$ can be determined 
by measuring the mean density of voids in the bulk (Fig.\ref{effective_scaling}). 
Similarly, in DOLB growth with desorption, $p_{\mathrm{eff}}$ is 
connected to the mean fraction of the removed material \cite{note02}.

\begin{figure}[tp]
\includegraphics[width=7.0cm]{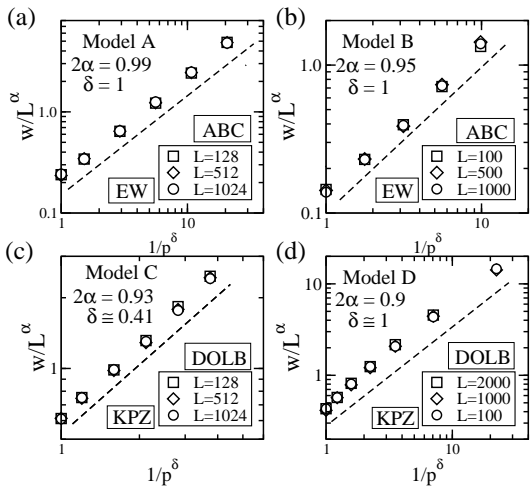}
\caption{\label{fig-1} 
Scaled widths at saturation $w$ vs the parameter
$1/p^{\delta}$: (a) and (b) are for {\it Models A} and {\it B}, respectively;
(c) and (d) are for {\it Models C} and {\it D}, respectively.
Reference lines have slope $1$.
Data are scaled with the $\alpha$ values shown.
}
\end{figure}

The analysis presented here explains scaling results of the following  
mixed-growth models in $(1+1)$ dimensions.   
{\it Model A} \cite{HA03,HMA01,KNV06}:  
component {\it 2} is RD with surface relaxation. 
{\it Model B} \cite{KNV06}:  component {\it 2} simulates a deposition 
of a sticky non-granular material of variable droplet size.  
{\it Model C} \cite{HA03,HA01,KNV06}: 
component {\it 2} is the NN sticking rule of BD. 
{\it Model D} \cite{KNV06,KNV04}:  component {\it 2} is a 
deposition of Poisson-random numbers to the local surface minima. 
{\it Models A} and {\it B} are ABC growths in the EW 
universality class, where $p_{\mathrm{eff}}=p$ and $\delta=1$ 
(Fig.\ref{fig-1}). {\it Models C} and {\it D} belong to the KPZ universality 
class. {\it Model C} is an example of DOLB growth with voids, 
with a 53.2\% void density in the bulk when $p=1$, and in this case 
$\delta \approx 0.41 <1$. {\it Model D} is a DOLB-type growth 
that produces a compact bulk but component {\it 2} is flux non-conserving, 
and here $\delta \approx 1$. 
Extensions of {\it Models A} and {\it C} to ($1+n$) 
dimensions \cite{HA03}, $n=2,3$, yield results that conform to 
our theoretical predictions of 
$p_{\mathrm{eff}} \approx p^\delta$ with $\delta \ne 1$ for 
mixing RD with DOLB processes, and $\delta =1$ for 
mixing RD with ABC processes. Additional examples
include cases when component {\it 2} is a restricted Kim-Kosterlitz 
solid-on-solid model \cite{Reis06}, and when 
it represents the Villain-Lai-Das Sarma universality class \cite{Reis04}.

The extension of the approach presented here to other competitive 
growth processes may provide a tool to understand the 
observed dynamics of surface growth. Realistic systems may involve 
many component-processes, some of which may be dominant. Within our 
formalism a departure point may be a generalization of Eq.(\ref{simplex}):
\begin{equation}
\label{g-simplex}
\Delta h(\vec{x},t) = \sum_k p_{\mathrm{eff}}^{(k)} \Delta h^{(k)} (\vec{x},t) \, ,
\end{equation}
where the summation is over contributing processes, and 
$ \Delta h^{(k)}$ is the column-height increment due to the $k$th process. 
In first approximation component-processes are not explicitly correlated. 
Each process is encountered with probability $p_k$, $\sum_k p_k=1$, and 
contributes to the growth with an effective probability $p_{\mathrm{eff}}^{(k)}$,   
$\sum_k p_{\mathrm{eff}}^{(k)} =1$. In the trivial case of all components being 
ABC type models with 
unit mean deposit height $p_{\mathrm{eff}}^{(k)}=p_k$. 
For a DOLB growth $p_{\mathrm{eff}}^{(k)}$ will have to be determined. 
Depending on the model, this can be done by analyzing the growth 
when process $k$ acts alone, and measuring either the mean bulk 
density or the mean fraction of the detached material or 
both \cite{note02}. Simplectic decompositions like the one proposed in 
Eq.(\ref{g-simplex}) have a long history of applications in many diverse fields.

In summary, the derived continuum equations and the resulting scaling 
show that model-dependent prefactors in universal scaling laws can be 
determined from bulk structures. This necessitates the distinction 
between the absorption-bulk-compact and the dense-or-lace-bulk growth 
processes in the analysis of competitive mixed-growth models.

\begin{acknowledgments}
We thank F. D. A. Aar\~{a}o Reis and H. E. Stanley for stimulating correspondence. 
This work is supported by NSF Grant DMR-0426488, and by CCS at MSU. 
It used resources of the NERSC Center, 
supported by the Office of Science of the US DoE 
under Contract No. DE-AC03-76SF00098.
\end{acknowledgments}


\begin{thebibliography}{99}
\bibitem{note02} The details will be given in extended publications.
\end{thebibliography}
\end{document}